\documentstyle[prb,aps,psfig]{revtex}
\newcommand \be{\begin{eqnarray}}
\newcommand \ee{\end{eqnarray}}
\begin{document}
\twocolumn[\hsize\textwidth\columnwidth\hsize
           \csname @twocolumnfalse\endcsname
\title{Bernoulli potential in type-I and weak type-II
superconductors: Surface dipole}
\author{P. Lipavsk\'y$^{1,2}$, K. Morawetz$^{3,4}$, 
J. Kol\'a\v cek$^1$, 
J. J. Mare{\v s}$^1$, E. H. Brandt$^5$, M. Schreiber$^3$}
\address{$^1$Institute of Physics, Academy of Sciences, 
Cukrovarnick\'a 10, 16253 Prague 6, Czech Republic\\
$^2$Faculty of Mathematics and Physics, Charles University,
Ke Karlovu 5, 12116 Prague 2, Czech Republic\\
$^3$Institute of Physics, Chemnitz University of Technology, 
09107 Chemnitz, Germany\\
$^4$Max-Planck-Institute for the Physics of Complex
Systems, Noethnitzer Str. 38, 01187 Dresden, Germany\\
$^5$Max-Planck-Institute for Metal Research,
         D-70506 Stuttgart, Germany}
\maketitle
\begin{abstract}
The Budd-Vannimenus theorem is modified to apply to  
superconductors in the Meissner state. The obtained identity 
links the surface value of the electrostatic potential to the 
density of free energy at the surface which allows one to 
evaluate the electrostatic potential observed via the capacitive 
pickup without the explicit solution of the charge profile. 
\end{abstract}
    \vskip2pc]

\section{Introduction}
The Hall voltage is commonly used to measure the concentration
of charge carriers in conductors. In superconductors this method is not 
applied, because of a missing theoretical support. In this 
paper we show that the Bernoulli potential, which is closely
related to the Hall voltage, can be used to the same end 
in superconductors.

Any ideal collision-less electric fluid should exhibit 
a finite Hall voltage, but superconductors seem to escape 
this theoretical conclusion. The zero Hall voltage was 
first reported by Kamerlingh Onnes and Hof\cite{KH14} 
already in 1914. Though later analyses showed that their 
samples were in the mixed state what obscures an 
interpretation of their experiment, the zero Hall voltage 
was confirmed anyway.\cite{L53}

From a theoretical point of view, it was clear that there 
has to be a voltage balancing the magnetic pressure which 
acts on electrons via the Lorentz force. The report of 
Kamerlingh Onnes and Hof thus stimulated various 
speculations about the missing Hall voltage, see the 
critical review by Lewis.\cite{L55} From various concepts 
proposed we mention the so-called contact potential -- a 
potential step at the interface -- which was expected to 
cancel the Hall voltage. Such a potential step might exist 
only if a charge dipole is formed at the interface of a 
superconductor. The surface dipole we discuss here is a 
similar concept.

The explanation of the zero Hall voltage turned out to be very
simple. By contacts one monitors the differences in the Gibbs
chemical potential, often called the electro-chemical 
potential.\cite{H66} 
The Gibbs potential is composed of three components: the 
electrostatic potential, the kinetic energy and the correlation 
energy. None of these components is constant in the presence 
of diamagnetic currents, but their sum is constant in equilibrium
in agreement with the observed zero Hall voltage.

To eliminate the kinetic and correlation energies, Hunt 
proposed to access the electrostatic potential with a 
contact-less method called the Kelvin capacitive 
pickup.\cite{H66} The first capacitive measurements 
appeared soon and they successfully proved the existence 
of a non-zero electrostatic potential.\cite{BK68,BM68} 

Both experiments\cite{BK68,BM68} were done at temperatures 
well below $T_c$, where the electrostatic potential has a 
simple form resembling the Bernoulli law,\cite{B37,L50,AW68}
\be
e\varphi=-{1\over 2}mv^2.
\label{b1}
\ee
To avoid confusion we note that the electrostatic potential in 
equilibrium superconductors is called the Bernoulli potential
for brevity, even if its actual form does not coincide with the 
Bernoulli law.

None of the early experimental data were sufficiently accurate 
to allow for a discussion of possible corrections to the 
plain Bernoulli potential (\ref{b1}). Nevertheless, the 
authors\cite{BK68,BM68} 
made some conclusions in this direction and we find it 
necessary to comment on them in more detail.

\subsection{Bok and Klein}
Bok and Klein\cite{BK68} claimed that their data agree with the 
plain Bernoulli potential (\ref{b1}). This conclusion has to be 
taken with reservations, however, because they measured the 
electrostatic potential as a function of the magnetic field $B$ at 
the surface. They evaluated the velocity ${\bf v}$ of the 
superconducting electrons (briefly called the condensate velocity) from 
the London condition $m{\bf v}=e{\bf A}$ and the exponential 
decay ${\bf A}={\bf A}_0{\rm e}^{-x/\lambda_0}$, using ${\bf B}=
\nabla\times{\bf A}$. At low temperatures, the London 
penetration depth depends on the density $n$ of pairable 
electrons, $\lambda^2_0=m/(e^2\mu_0n)$, therefore their 
experimental result can be expressed in terms of the 
magnetic pressure, $en\varphi=- B^2/(2\mu_0)$. 

According to the above arguments, the relation of the 
electrostatic potential to the magnetic pressure seems to be
a consequence of the plain Bernoulli potential and the London
theory. As we show below, the relation to the magnetic 
pressure is very general and holds also under conditions when 
neither the plain Bernoulli potential nor the London theory 
applies. 

In fact, Bok and Klein measured on indium which, being a 
type-I superconductor, is not fully covered by the London 
theory. Moreover, they swept the magnetic field from zero to 
the critical value, while the plain Bernoulli potential and 
the simple form of the London theory used above are 
restricted to low magnetic fields. At high fields, the 
condensate density at the surface is suppressed what results
in $v^4$ and higher-order contributions to the electrostatic
potential.

Briefly, Bok and Klein have observed the magnetic pressure.
But the link of their experiment to the Bernoulli potential 
(\ref{b1}) has to be taken with caution.

\subsection{Brown and Morris}
Brown and Morris have used a different setup which allowed
them to achieve a much higher precision.\cite{BM68} They did
not control the magnetic field but the current in a thin wire. 
This current was scaled with the critical current. They 
announced in 1968 that their data reveal about 20\% deviations 
from the screened Bernoulli potential, 
\be
e\varphi=-{n_s\over n}{1\over 2}mv^2,
\label{b2}
\ee
discussed in more detail below.

It should be noted that Brown and Morris expected deviations 
which were predicted from the BCS theory in the same year. 
Adkins and Waldram had studied the electrostatic potential 
from changes of the BCS gap due to a current and they 
recovered the plain Bernoulli potential for zero temperature, 
while for finite temperatures they indicated a presence of 
additional contributions.\cite{AW68} They were not, however, 
capable to derive these contributions in an explicit form or to 
estimate their amplitudes.

Some corrections to the Bernoulli potential (\ref{b1}) were
derived already before the BCS studies. Historically the 
first is the theory of Sorokin\cite{S49} from 1949 which 
covers the majority of effects recovered later. Although this
paper is mentioned by London,\cite{L50} later it became 
forgotten. In 1964 van~Vijfeijken and Staas\cite{VS64} took
into account that the electrostatic field acts on normal
electrons and arrived at the so-called quasiparticle screening. 

The quasiparticle screening is represented by the fraction 
of superconducting electrons $n_s/n$ by which eq. 
(\ref{b2}) differs from the simple Bernoulli law (\ref{b1}). 
In the experiment of Brown and Morris it accounts 
for  6\% of the observed potential. In spite of its small
magnitude, the quasiparticle screening is important with 
respect to the concept of the magnetic pressure. At finite 
temperatures one has to take into account that the London 
penetration depth also depends on the density of condensate, 
$\lambda^2=m/(e^2\mu_0n_s)=\lambda_0^2n/n_s$. 
Combining the screened Bernoulli potential (\ref{b2}) with the 
condensate velocity $v$ found from the London theory, one 
finds that the electrostatic potential is temperature 
independent, i.e., it is given by the magnetic pressure with 
the density $n$ of pairable electrons. Despite this importance 
of the quasiparticle screening, it should be noted, however, 
that the quasiparticle screening is not responsible for the 
20\% deviations announced by Brown and Morris
and discussed above.

Corrections capable to explain the observed potential were 
first obtained by Rickayzen,\cite{R69} who showed that the 
electrostatic potential includes a pairing contribution, 
\be
e\varphi=-{\partial n_s\over\partial n}
{1\over 2}mv^2=-{n_s\over n}{1\over 2}mv^2-4{n_n\over n}
{\partial\ln T_c\over\partial\ln n}{1\over 2}mv^2.
\label{b3}
\ee 
The pairing term dominates close to the critical temperature 
$T_c$, because $n_s=n(1-t^4)$, with $t=T/T_c$, while $n_n=
n-n_s=nt^4$. 

According to (\ref{b3}), from $\varphi$ close to $T_c$ one 
may deduce the density dependence of $T_c$, which would be   
very attractive with respect to designing new materials. Indeed,
this important material property is otherwise deducible only 
from measurements applying a hydrostatic pressure or 
adding impurities to crystals. Unlike the later methods, 
a measurement of the Hall voltage or, respectively, the 
Bernoulli potential does not affect the 
electronic bands, the phonon spectrum or the electron-phonon 
interaction, therefore it offers a uniquely clear information 
about the material.

All expectations were chilled by the next paper of Morris and 
Brown.\cite{MB71} They admitted that deviations announced in 
the first paper were due an incorrect estimate of the critical 
current and presented new highly accurate data for a wide 
range of temperatures. The observed electrostatic potential 
is perfectly equal to the magnetic pressure and exhibits no 
pairing contribution. They reported that this behavior is 
common to both type-I and weak type-II superconductors and 
for the magnetic field up to the critical value.

\subsection{Surface dipole}
The disagreement between theory and experiment remained 
unexplained for a long time and the question of the charge 
transfer in superconductors was left aside till the discovery of 
the high-$T_c$ materials. For these layered materials it was 
predicted\cite{M89,KK92} that  the superconducting transition 
induces a charge transfer from CuO$_2$ planes to charge 
reservoirs. This transfer caused merely by the pairing 
mechanism has been confirmed by bulk- oriented experiments 
like the positron annihilation,\cite{positron} the \mbox{$x$-ray} 
absorption spectroscopy,\cite{xray} and the nuclear magnetic 
resonance.\cite{KNM01}

Apparently, there are two groups of contradictory experimental
results. The pairing contribution is absent in the surface 
potential but a charge transfer is observed at internal 
interfaces in the bulk. 

As it was indicated recently,\cite{LKMM02} there is a charge 
transfer at the surface which is the interface of superconductor 
and vacuum. This transfer forms a surface dipole which causes 
a step $\varphi_\delta$ in the electrostatic potential. The value 
of the potential step has been evaluated from the 
Budd-Vannimenus theorem\cite{BV73} 
\be
e\varphi_\delta=n{\partial\over\partial n}{f_{\rm el}\over n},
\ee 
where $f_{\rm el}$ is the electronic part of the free energy 
density (it does not include the electrostatic and magnetic 
parts). Rickayzen has obtained formula (\ref{b3}) from the 
general stability condition $e\varphi=-{\partial f_{\rm el}\over
\partial n}$ and the free energy $f_{\rm el}=n_s{1\over 2}mv^2$. 
Using the same free energy one finds that the potential at the 
surface, $\varphi_0=\varphi(0)+\varphi_\delta$, equals the 
screened Bernoulli potential (\ref{b2}). The surface dipole 
$\varphi_\delta$ thus explains the observed magnetic pressure.

In spite of the agreement between observed and theoretically
derived voltage one should be reserved about claims that the
theory correctly describes the profile of the electrostatic 
potential in superconductors. Arguments against the theory are 
similar to those already raised in relation to the interpretation of 
the measurement of Bok and Klein.

First, the measurement of Morris and Brown explores the entire 
range of magnetic fields from low up to critical values. The free 
energy employed by Rickayzen, however, applies in the limit of 
low magnetic fields only. 

Second, materials studied by Morris and Brown are type-I 
and weak type-II materials so that their behavior is not fully 
covered by the London theory. As we have shown 
recently,\cite{LMKMBSa03} even for low magnetic fields the 
electrostatic potential depends on the Ginzburg-Landau (GL) 
parameter $\kappa$. Rickayzen's formula is recovered for 
the extreme type-II superconductor $\kappa\to\infty$. For 
measured materials with $\kappa\approx 1/\sqrt{2}$ the 
potential $\varphi(0)$ at the surface is reduced by a factor 
$1/3$ compared to Rickayzen's formula.

Third, the surface dipole $\varphi_\delta$ derived from the free 
energy that covers only the low-field perturbation in the London
approximation has the same shortcomings as Rickayzen's formula.
Apparently, $\varphi(0)$ obtained from Rickayzen's formula 
can be far from the correct value and the same applies to 
$\varphi_\delta$. Since the sum agrees with the 
experimental result, one can see that eventual errors tend to
compensate each other in the resulting electrostatic potential.
In this sense, the surface dipole $\varphi_\delta$ is consistent 
with the internal potential $\varphi$, since both are evaluated
using the same free energy.

\subsection{Plan of the paper}
As demonstrated for Rickayzen's theory, the internal 
electrostatic potential $\varphi$ and the surface dipole 
$\varphi_\delta$ ought to be derived from the same free 
energy. We have shown\cite{LMKMBSa03} that for type-I 
and weak type-II superconductors, the GL theory yields the 
internal electrostatic potential which is quite different from the 
one predicted by Rickayzen's formula. In this paper we derive 
the surface dipole within the GL theory. 

As in Ref.~\onlinecite{LKMM02} we use the Budd-Vannimenus 
theorem. Here we employ this identity within Bardeen's 
extension\cite{B55} of the GL theory. 
Bardeen's extension offers two advantages. First, it naturally 
interpolates between the GL theory close to $T_c$ and the 
London theory at low temperatures. Second, it uses material 
parameters of the Gorter-Casimir two-fluid model, which have 
a transparent density dependence. In contrast, the parameters 
of the original GL theory are introduced in the limit 
$T\to T_c$ and $T$ is replaced by $T_c$ wherever possible.
Since $T$ is an independent thermodynamic variable while 
${\partial T_c\over\partial n}\ne 0$, one has to be careful 
when taking density derivatives. To evaluate the density 
dependence of the GL parameters, one has to recall the 
microscopic theory of Gorkov and take the density derivatives 
of the corresponding parameters before the limit $T\to T_c$ is 
applied.

The plan of the paper is as follows. In Sec.~II we assume 
temperatures close to $T_c$ and an infinitesimally weak 
magnetic field. In this limit we derive a modification of the 
Budd-Vannimenus theorem which takes into account a 
non-zero charge density near the surface. In Sec.~III we 
discuss a general system in the Meissner state and show that 
the formula for the surface dipole derived in Sec.~II applies to 
any temperature and magnetic field below critical values. 
Section~IV includes a summary.

\section{Surface dipole}\label{S2}
Let us first estimate the thickness of the surface dipole from
thermodynamic considerations. The pairing correlation is 
weaker on the surface than in the bulk, what results in forces 
pulling the Cooper pairs inside. Such forces are always 
balanced by the electrostatic field. The full understanding of 
this effect will require microscopic studies which are not yet 
feasible. From the BCS studies it is known, however, that close 
to the surface on the scale of the BCS coherence length 
$\xi_0$, the gap profile differs from the value given by the GL 
theory.\cite{G66} We thus expect that the surface dipole is 
somehow linked to this `microscopic' modulation of the gap 
profile.

To introduce the surface dipole on an intuitive level, let us 
assume that the system is close to the critical temperature. 
In this regime, the London penetration depth $\lambda$ and 
the GL coherence length $\xi$ are much larger than the BCS 
coherence length $\xi_0$. Since the electrostatic potential 
induced by the diamagnetic current extends on scales of 
$\lambda$ and $\xi$ from the surface,\cite{LMKMBSa03} the 
surface dipole is very narrow on these scales.

\begin{figure}[h]  
\psfig{figure=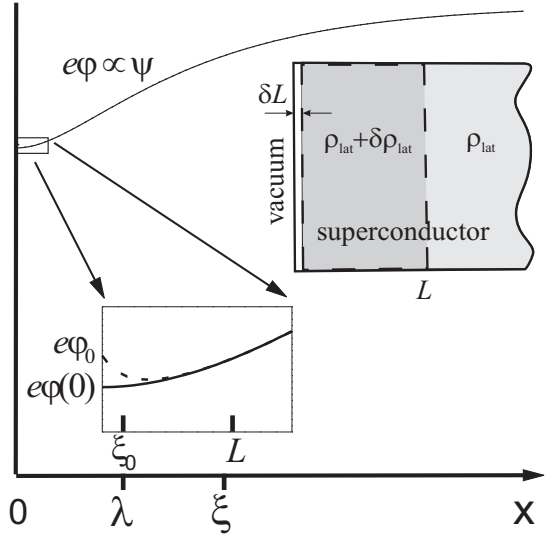,width=8cm}
\vskip 2pt
\caption{Electrostatic potential at the surface. On the scale 
of the London penetration depth $\lambda$ and the GL 
coherence length $\xi$, the electrostatic potential $\varphi$ 
is linear in the GL wave function $\psi$. The schematic 
short-dashed line in the expanded detail shows the deviation 
of the potential from the GL value on the scale of the 
BCS coherence length $\xi_0$. The surface potential
$\varphi_0$ is the experimentally observed value, while 
$\varphi(0)$ is the internal GL potential extrapolated to the 
surface. In the insert, we show the virtual compression of the 
crystal lattice on the scale $L$, as it is employed in the 
Budd-Vannimenus theorem. The compression removes the 
lattice charge from the infinitesimal layer $\delta L$ and 
correspondingly increases the charge density in the region
framed by the long-dashed line.}
\label{f1}
\end{figure}

We can then define an intermediate scale $L$ such that 
$\xi_0\ll L\ll\xi,\lambda$, as sketched in Fig.~\ref{f1}. On 
the scale $L$, the GL wave function changes only negligibly, 
i.e., $\psi(x) \approx\psi(x\to 0)\equiv\psi(0)$ for $0<x<L$. 
We note that the GL boundary condition $\partial_x\psi=0$ 
supports the slow change of $\psi$ close to the surface.

As shown by de~Gennes\cite{G66}, the GL wave function is 
linearly proportional to the BCS gap, except for the surface 
region on the scale of the BCS coherence length $\xi_0$. 
Following microscopic theories giving the electrostatic potential 
in terms of the BCS gap,\cite{M89,KK92,H75} we expect the 
electrostatic potential to have similar features, see 
Fig.~\ref{f1}. 

For $\xi_0\ll x<L$, the potential is well described by the 
GL value $\varphi(x)$. Since $L\ll\lambda,\xi$, the GL 
prediction of the electrostatic potential changes only 
negligibly in this region and it is convenient to 
introduce the extrapolated value $\varphi(x)\approx\varphi
(x\to 0)\equiv\varphi(0)$. The extrapolation of the GL 
potential towards the surface, $\varphi(0)$, has to be 
distinguished from the true surface potential $\varphi_0$. 
The difference $\varphi_\delta=\varphi_0-\varphi(0)$ is 
caused by the surface dipole we aim to evaluate.

\subsection{Budd-Vannimenus theorem at intermediate scale}
Close to the critical temperature, we can take the intermediate 
region $x\sim L$ as a homogeneous `bulk' and follow the idea 
of Budd and Vannimenus.\cite{BV73,KW96} Let us assume 
a virtual compression of the crystal lattice such that the 
background or the lattice charge density is removed from the 
surface layer of an infinitesimal width $\delta L$. The 
perturbation of the lattice charge density in the infinitesimal 
layer $0<x<\delta L$ is $\delta\rho_{\rm lat}=-\rho_{\rm lat}$. 
The compression leads to an increase of the charge density in 
the layer $\delta L<x<L$, where $\delta\rho_{\rm lat}=
\rho_{\rm lat}\delta L/L$ is selected to conserve the total 
charge. 

Now we recall the basic idea of the Budd-Vannimenus 
theorem. The lattice charge enters the jellium model of metals 
as an external parameter. If one changes this external 
parameter, the situation corresponds to doing work on the 
system, 
\be
\delta{\cal W}= S\int d x\delta\rho_{\rm lat}{\partial 
f\over\partial\rho_{\rm lat}}=S\int d x \delta\rho_{\rm lat} 
\varphi,
\ee 
where $f$ is the density of the free energy including
the electrostatic interaction, and $S$ is the sample area. 
According to the Feynman-Hellmann theorem, the change of 
the electrostatic potential does not contribute to the work up to
the first order in $\delta\rho_{\rm lat}$. Now we can proceed
with the algebra. We split the integral into three parts,
$(0,\delta L)$, $(\delta L,\xi_0)$ and $(\xi_0,L)$. Since 
$\delta L$ is an infinitesimal displacement, the potential in the 
layer $0<x<\delta L$ can be replaced by the surface value 
$\varphi_0$. The surface region $\delta L<x<\xi_0$ gives a 
negligible contribution of the order of $\xi_0/L$. In the 
remaining bulk region $\xi_0<x<L$, the electrostatic potential 
is nearly constant and equals $\varphi(0)$. The work thus 
reads 
\be
\delta {\cal W}=S\delta L\rho_{\rm lat}(\varphi(0)-\varphi_0).
\label{dw}
\ee 

The work increases the free energy ${\cal F}$ of the system
\be
\delta{\cal W}=\delta{\cal F}=-{\partial (f_{\rm el}SL) 
\over\partial (S L)}S\delta L=\left(-f_{\rm el}+
n{\partial f_{\rm el}\over\partial n}\right)S\delta L,
\label{df}
\ee 
where $f_{\rm el}$ is the spatial density of the electronic 
free energy. Note that only the change of the electronic part
in the `bulk' is assumed. The surface energy does not change as 
the surface is merely shifted. The magnetic free energy 
changes negligibly, because the number of electrons in the 
layer $L$ is not changed by the deformation. Since the 
condensate velocity changes on the scale $\lambda\gg L$, 
changes of the screening current vanish in the limit 
$L/\lambda\to 0$.

The first term in (\ref{df}) results from the reduced volume, 
$SL\to S(L-\delta L)$, and the second one from the 
corresponding increase of the electron density, $n\to 
n(1+\delta L/L)$. Equating (\ref{dw}) and (\ref{df}) we obtain 
a modification of the Budd-Vannimenus theorem,
\begin{equation}
\rho_{\rm lat}\varphi_\delta\equiv
\rho_{\rm lat}\left(\varphi_0-\varphi(0)\right)=
f_{\rm el}-n{\partial f_{\rm el}\over\partial n}.
\label{e1}
\end{equation} 
It describes the step of the potential at the surface due to the 
surface dipole in terms of the free energy. This relation is the 
main result of the present paper. In the remaining part we 
demonstrate how (\ref{e1}) can be used and show that it 
applies also at low temperatures, where the intermediate 
region cannot be defined. 

Formula (\ref{e1}) differs from the original Budd-Vannimenus 
(BV) theorem in four points. First, in the original BV 
theorem one evaluates the total potential step at the surface.
Its values are of the order of Volts. Here we evaluate only 
the change of the potential step which appears as the system 
goes superconductive. The typical magnitude of this change is of
the order of nano-Volts. Second, in the BV theorem the surface 
potential is related to the potential $\varphi_\infty$ deep 
in the bulk. In (\ref{e1}), the extrapolation of the internal 
potential towards the surface, $\varphi(0)$, appears instead. 
Third, in order to cover systems at finite temperatures, we 
use the free energy instead of the ground-state energy. 
Fourth, within the original Budd-Vannimenus approach, the 
electron charge density and the lattice charge density have 
to be equal because of the charge neutrality. In our approach, 
the density of electronic charge differs locally from the 
lattice charge density, $en\ne -\rho_{\rm lat}$, due to the 
charge transfer induced by the magnetic 
fields.\cite{LMKMBSa03}

In the limit of extreme type-II superconductors and weak
magnetic fields one can compare formula (\ref{e1}) with
the surface dipole evaluated in Ref.~\onlinecite{LKMM02}.
In this limit, the free energy simplifies to $f_{\rm el}=
f_\infty-n_s{1\over 2}mv^2$. The second term provides the
current induced changes of the surface dipole derived in
Ref.~\onlinecite{LKMM02}. Due to the bulk free energy 
$f_\infty$ one finds from (\ref{e1}) a finite potential 
step also in the absence of diamagnetic currents. 

For the purpose of the plot in Fig.~\ref{f1}, we have 
used a small magnitude of the potential step at the surface
(the short-dashed line is rather close to the full line). 
In reality, even the current induced part of the step 
$\varphi_\delta$ might achieve a magnitude much larger 
than the internal potential $\varphi(0)$. A simple estimate 
of surface step for the experiment of Morris and 
Brown\cite{MB71} ranges from few percent for $T=0.6\,T_c$ 
to about thirty times of the observed potential at 
$T=0.97\,T_c$, see Ref.~\onlinecite{LKMM02}. Since we 
assume the limit $T\to T_c$, the large values are more 
appropriate.

\subsection{Application within the GL theory}
Now we demonstrate how the relation (\ref{e1}) can be used 
within the GL theory. To this end we introduce the GL free 
energy 
\begin{equation}
f_{\rm el}={1\over 2m^*}\left|\left(-i\hbar\nabla-
e^*{\bf A}\right)\psi\right|^2+f_{\rm cond},
\label{e2}
\end{equation}
where $\psi$ is the GL wave function, ${\bf A}$ is the vector 
potential, $m^*=2m$ and $e^*=2e$ are the mass and the 
charge of the Cooper pair, and $f_{\rm cond}$ is the free 
energy of the condensate. It can be either the Gorter-Casimir 
free energy 
\be
f_{\rm cond}=-{1\over 4}\gamma T_c^2{2\over n}
|\psi|^2-{1\over 2}\gamma T^2\sqrt{1-{2\over n}|\psi|^2}
\ee 
used by Bardeen\cite{B55}  or the GL free energy 
\be
f_{\rm cond}=
\alpha|\psi|^2+{1\over 2}\beta|\psi|^4.
\ee 
The GL parameters $\alpha=\gamma(T_c^2-T^2)/2n$ and 
$\beta=\gamma T^2/n^2$ depend on the temperature $T$ and 
the electron density $n=n_n+2|\psi|^2$, where $n_n$ is the 
density of normal electrons. Finally, we add the 
electromagnetic energy so that the free energy reads
\begin{equation}
f=f_{\rm el}+\varphi(\rho_{\rm lat}+en)-{\epsilon_0\over 2}E^2+
{1\over 2\mu_0}B^2,
\label{e3}
\end{equation}
with the magnetic field ${\bf B}=\nabla\times {\bf A}$ and the 
electric field ${\bf E}=-\nabla\varphi$. 

Variations of the free energy with respect to its independent 
variables ${\bf A},\varphi,\psi,n_n$ yield the equations of motion in 
Lagrange's form
\begin{equation}
-\nabla{\partial f\over\partial\nabla\nu}+{\partial f\over\partial
\nu}=0.
\label{e4}
\end{equation}
For $\nu=A$ the variational condition (\ref{e4}) yields the 
Ampere-Maxwell equation, for $\nu=\varphi$ the Poisson 
equation, for $\nu=\psi$ the GL equation, and for $\nu=n_n$ 
the condition of zero dissipation,
\begin{equation}
e\varphi=-{\partial f_{\rm el}\over\partial n}.
\label{e5}
\end{equation}
This condition allows one to evaluate the electrostatic potential 
in the bulk of the superconductor.\cite{LKMB02} Of course, one can add any constant to the electrostatic potential. 

Formula (\ref{e5}) does not cover the surface dipole on the
scale $\xi_0$, therefore at the surface it provides the 
extrapolated bulk value $\varphi(0)$. We can thus use (\ref{e5}) to rearrange the Budd-Vannimenus theorem (\ref{e1}) as
\begin{equation}
\rho_{\rm lat}\varphi_0=f_{\rm el}+\varphi(0)
\left(\rho_{\rm lat}+en\right).
\label{e6}
\end{equation}
Now all terms on the right hand side are explicit quantities 
which one obtains within the GL theory extended by the 
electrostatic interaction.\cite{LKMB02}

\subsection{Convenient approximation}
In customary GL treatments, the electrostatic potential and the 
corresponding charge transfer are omitted. For magnetic 
properties this approximation works very well, since the relative 
charge deviation, $\left(\rho_{\rm lat}+en\right)/\rho_{\rm lat}$, 
is typically of the order of $10^{-10}$, leading to comparably 
small corrections in the GL equation. With the same accuracy 
one obtains the electronic free energy $f_{\rm el}$. Therefore it 
is possible to evaluate the surface potential using the 
approximation
\begin{equation}
\varphi_0\approx -{f_{\rm el}\over en_\infty}
\label{e7}
\end{equation}
which follows from (\ref{e6}) if terms proportional to 
$\left(\rho_{\rm lat}+en\right)/\rho_{\rm lat}$ are neglected. 
By $n_\infty$ we have denoted the asymptotic value of $n$ 
deep in the bulk, i.e. the density of pairable electrons, 
$\rho_{\rm lat}=-en_\infty$.

Within approximation (\ref{e7}) one does not have to 
evaluate the potential profile and the related charge inside the 
superconductor. This is advantageous, in particular for systems
of unknown material parameters $\partial T_c/\partial n$ and
$\partial\gamma/\partial n$.

Turning the argument around, from (\ref{e7}) one can see that
the electrostatic potential at the surface cannot be used to
measure the material parameters $\partial T_c/\partial n$ and
$\partial\gamma/\partial n$. This fact is already known from 
the experiment of Morris and Brown.\cite{MB71}

\section{Magnetic pressure}
So far we have discussed systems close to the 
critical temperature, when the validity conditions of the GL 
theory are well satisfied. In many cases, however, the GL 
theory is used beyond the limits of its nominal applicability.
In these cases the GL coherence length $\xi$ and/or 
the London penetration depth $\lambda$ are comparable to, or 
even shorter than the BCS coherence length $\xi_0$ so that
the intermediate scale $L$ cannot be introduced.

It is possible, however, to follow the original 
formulation of Budd and Vannimenus and define $L$ as the
sample thickness, i.e. $L\gg\xi,\lambda$. In this case it is
necessary to account for the energy of the magnetic field, 
since the infinitesimal compression $\delta L$ shifts the 
screening layer inwards into the superconductor. 

\subsection{Budd-Vannimenus theorem}
For $L\gg\xi,\lambda$, the charge removed from the surface 
is placed deep in the bulk (the region of the scale of 
$\xi,\lambda$ gives a negligible contribution) so that the work 
on the charge reads
\be
\delta{\cal W}=S\delta L(\varphi_\infty-\varphi_0)
\rho_{\rm lat}.
\ee 
Compared to the previous treatment
we have merely replaced the potential close to the surface by
the value deep in the bulk. Similarly, the electronic part of the 
total free energy changes by 
\be
\delta{\cal F}_{\rm el}=
\left(-f_{\rm el}^\infty+n{\partial f_{\rm el}^\infty\over\partial n}
\right)S\delta L.
\ee 
Finally, the shift of the screening layer by $\delta L$ changes 
the magnetic energy by an amount
$\delta {\cal F}_B=S\delta LB_0^2/2\mu_0$, given by the 
magnetic pressure. Here $B_0$ is the value of the magnetic
field at the surface.

From $\delta{\cal W}=\delta{\cal F}_{\rm el}+\delta{\cal F}_B$
follows
\begin{equation}
\rho_{\rm lat}(\varphi_0-\varphi_\infty)={B_0^2\over 2\mu_0}+
f_{\rm el}^\infty+n_\infty{\partial f_{\rm el}^\infty\over\partial n}.
\label{e8}
\end{equation}
As $L$ is the thickness of the sample one can use the
charge neutrality, $\rho_{\rm lat}=-en_\infty$. Since the value 
of the potential deep in the sample is given by condition 
(\ref{e5}), using $e\varphi_\infty=-{\partial f_{\rm el}^\infty
\over\partial n}$ from (\ref{e8}) one obtains 
\begin{equation}
\varphi_0=-{B_0^2\over 2\mu_0en_\infty}-
{f_{\rm el}^\infty\over en_\infty}.
\label{e8a}
\end{equation}
The electrostatic potential observed at the surface is thus
given by the magnetic pressure as observed by Morris and
Brown.\cite{MB71}

Note that deriving formula (\ref{e8a}) we have not used 
many assumptions about the system. The condition of zero 
dissipation (\ref{e5}) is a general thermodynamic relation. 
The Budd-Vannimenus relation (\ref{e8}), however, is limited 
to systems with a homogeneous jelly-like background 
charge. This approximation is acceptable for conventional
superconductors, where characteristic scales $\xi_{\rm BCS},
\xi,\lambda$ are much larger than the elementary cell of the
crystal. The applicability is questionable for the high-$T_c$
materials which due to the layered structure and a short 
coherence length are far from the jellium model.

As noticed already by Bok and Klein, \cite{BK68} there is a 
simple argument for the formula like (\ref{e8a}). If one assumes
a slab with magnetic fields $B_L$ and $B_R$ on the left/right
sides, the voltage difference gives 
\be
\rho_{\rm lat}(\varphi_0^L-
\varphi_0^R)={1\over 2\mu_0}(B_L^2-B_R^2).
\ee 
The left hand side of this relation represents the electrostatic 
force (per unit area) on the lattice, 
\be
F_{\rm elst}=\int_L^R dx\,E\rho_{\rm lat}=
\rho_{\rm lat}(\varphi_0^L-\varphi_0^R).
\ee 
The right hand side is the Lorentz force $F_{\rm Lor}=BJ$ with 
the mean magnetic field $B={1\over 2}(B_L+B_R)$ and the net 
current $J=\int_L^R dx\,j$ given by Ampere's rule, $B_L-
B_R=\mu_0 J$. Since the electrostatic field provides the only 
mechanism by which the force is passed from the electrons to 
the lattice, the two forces have to be equal, $F_{\rm Lor}=
F_{\rm elst}$. This argument was, however, overlooked in the 
later studies.

\subsection{Test of the surface relation}
The Budd-Vannimenus theorem provides the electrostatic 
potential (\ref{e8a}) in terms of the magnetic field with no 
regards to the actual potential inside the superconductor. To 
link formula (\ref{e8a}) with the more intuitive derivation from 
Sec.~\ref{S2}, we show that (\ref{e6}) results in the surface 
potential (\ref{e8a}) for any temperature. 

For the assumed geometry, the GL equation has an integral 
of motion, see Bardeen.\cite{B55} This integral can be 
obtained quite generally by the Legendre transformation of 
the free energy, 
\begin{equation}
g=f-\sum_\nu{\partial f\over\partial\nabla\nu}
\nabla\nu.
\label{e8b}
\end{equation}
Indeed, if the fields $\nu$ obey the equations of motion 
(\ref{e4}), the gradient $\nabla g=0$ vanishes, i.e. 
$g={\rm const}$. Deep in the bulk all gradients vanish, 
therefore $g=f_{\rm el}^\infty$.

From equations (\ref{e2}-\ref{e3}) one finds
\begin{equation}
\sum_\nu{\partial f\over\partial\nabla\nu}\nabla\nu=
{\hbar^2\over m^*}|\nabla\psi|^2-\epsilon_0E^2+
{B^2\over\mu_0}.
\label{e8c}
\end{equation}
With the help of (\ref{e8b}-\ref{e8c}) and definition (\ref{e3}), 
one can express the electronic free energy as
\begin{equation}
f_{\rm el}=f_{\rm el}^\infty+{\hbar^2\over m^*}|\nabla\psi|^2-
{\epsilon_0E^2\over 2}
+{B^2\over 2\mu_0}-\varphi(\rho_{\rm lat}+en).
\label{e9}
\end{equation}
At the surface, the GL boundary condition demands that 
$\nabla\psi=0$ what implies $E=0$. The free energy at the 
surface thus reads
\begin{equation}
f_{\rm el}=f_{\rm el}^\infty+{B^2\over 2\mu_0}-\varphi(0)
(\rho_{\rm lat}+en).
\label{e10}
\end{equation}
From (\ref{e10}) and the surface relation (\ref{e6}) it follows that $\rho_{\rm lat}\varphi_0=B^2/(2\mu_0)+f_{\rm el}^\infty$. This 
value is identical to (\ref{e8a}).

Apparently, we can reverse the procedure. Starting from the
general Budd-Vannimenus relation (\ref{e8a}) and the general
integral of motion (\ref{e8b}), we can derive the surface 
relation (\ref{e6}). Accordingly, the surface relation holds for
any temperature, provided that the free energy is a functional
of the GL wave function and its first derivative only, $f\equiv 
f[\psi,\nabla\psi]$. This functional can be an arbitrary one.

Perhaps we should explain why we have derived the surface
dipole from the Budd-Vannimenus theorem on the intermediate
scale, although the more general derivation from the integral 
of motion is available. There are two reasons. First, the 
intermediate scale provides at least a qualitative picture of 
the potential in the vicinity of the surface. This picture might 
be helpful if measurements sensitive to layers close to the 
surface will be designed.

Second, within the intermediate scale the surface dipole is 
treated as a property of the superconducting condensate, 
what encourages us to hope that formula (\ref{e6}) or its 
approximation (\ref{e7}) can be used to obtain the surface 
potential also for cases when the magnetic field has a 
component perpendicular to the surface. In particular, we 
expect that it will be applicable to the superconductors 
in the mixed state, especially to evaluate the electric field 
generated by vortices penetrating the surface.\cite{B96}

\section{Conclusions}
In conclusion, the Budd-Vannimenus theorem was modified so 
that it is applicable to the surface of a superconductor. It allows 
one to evaluate the electrostatic potential on the surface 
from the free energy and the bulk electrostatic potential nearby. 
Formula (\ref{e7}) offers the approximation of the surface 
potential from the free energy without the actual knowledge 
of the bulk potential.

For plain surfaces we have recovered the experimentally 
established fact that the electrostatic potential equals
the magnetic pressure divided by the density of pairable
electrons. This experimental law was confirmed also for 
type-I and weak type-II superconductors, while the  
previous theoretical treatments were restricted to weak 
magnetic fields and extreme type-II superconductors.
The presented theory is free of these limitations. 

It was shown that thermodynamic corrections do not influence 
the surface electrostatic potential, measurable e.g. via contactless
capacitive pickup. Consequently, contrary to earlier expectations, 
the density dependence of the critical temperature cannot be estimated 
in this way. On the other hand, the relation between the
surface electrostatic potential and the magnetic pressure 
shows, that such a measurement allows one to determine the density of 
charge carriers without knowledge of any other material parameters.

In this paper we have derived only the amplitude of the 
potential step. The detailed profile of the electrostatic potential 
including its modulation at the surface can be obtained by a 
microscopic approach like the Bogoljubov-de~Gennes theory 
extended recently to cover the electrostatic 
phenomena.\cite{K01,MK02,MK03,MK03L} 
For microscopic calculations, the Budd-Vannimenus
theorem can serve as a test of accuracy of the numerical 
procedure, similarly as it is used in the theory of metal 
surfaces.

\acknowledgements
This work was supported by M\v{S}MT program Kontakt 
ME601 and GA\v{C}R 202/03/0410, GAAV A1010312 grants. 
The European (ESF) program VORTEX is also acknowledged.


\begin{references}

\bibitem{KH14}
H.~Kamerlingh Onnes and K.~Hof, 
Leiden. Comm.  {\bf 142}b (1914).

\bibitem{L53}
H.~W.~Lewis, Phys. Rev. {\bf 92}, 1149 (1953).

\bibitem{L55}
H.~W.~Lewis, Phys. Rev. {\bf 100}, 641 (1955).

\bibitem{H66}
T.~K.~Hunt, Phys. Lett. {\bf 22}, 42 (1966).

\bibitem{BK68}
J.~Bok and J.~Klein,
Phys. Rev. Lett. {\bf 20}, 660 (1968).

\bibitem{BM68}
J.~B.~Brown and T.~D.~Morris,
Proc. 11th Int. Conf. Low. Temp. Phys.,
Vol.~2, 768 (St. Andrews, 1968).

\bibitem{B37}
F.~Bopp, Z. Phys. {\bf 107}, 623 (1937).

\bibitem{L50}
F.~London, {\em Superfluids}
(Willey, New York, 1950), Vol.~I, Sec.~8.

\bibitem{AW68}
C.~J.~Adkins and J.~R.~Waldram, 
Phys. Rev. Lett. {\bf 21}, 76 (1968).

\bibitem{S49}
V.~S.~Sorokin, JETP {\bf 19}, 553 (1949).

\bibitem{VS64}
A.~G.~van~Vijfeijken and F.~S.~Staas,
Phys. Lett. {\bf 12}, 175 (1964).

\bibitem{R69}
G.~Rickayzen, J. Phys. C {\bf 2}, 1334 (1969).

\bibitem{MB71}
T.~D.~Morris and J.~B.~Brown,
Physica {\bf 55}, 760 (1971).

\bibitem{M89}
D.~van~der~Marel,
Physica C {\bf 165}, 35 (1990).

\bibitem{KK92}
D.~I.~Khomskii and F.~V.~Kusmartsev,
Phys. Rev. B {\bf 46}, 14245 (1992).

\bibitem{positron}
Y.~C.~Jean {\em et al.}, 
Phys. Rev. Lett. {\bf 64}, 1593 (1990);
A.~Bharati {\em et al.},
{\bf 42}, 10199 (1990);
P.~K.~Pujari {\em et al.},
Phys. Rev. B {\bf 50}, 3438 (1994);
U.~De {\em et al.}, 
Phys. Rev. B {\bf 62}, 14519 (2000).

\bibitem{xray}
Y.~Hirai {\em et al.},
Phys. Rev. B {\bf 45}, 2573 (1992);
N.~L.~Saini {\em et al.},
Phys. Rev. B {\bf 52}, 6219 (1995);
B.~R.~Sekhar {\em et al.},
Phys. Rev. B {\bf 56}, 14809 (1997).

\bibitem{KNM01}
K.~I.~~Kumagai, K.~Nozaki and Y.~Matsuda,
Phys. Rev. B {\bf 63}, 144502 (2001).

\bibitem{LKMM02}
P.~Lipavsk{\'y}, J.~Kol{\'a}{\v c}ek, J.~J.~Mare{\v s} and
K.~Morawetz,
Phys. Rev. B {\bf 65}, 012507 (2002).

\bibitem{BV73}
H.~F.~Budd and J.~Vannimenus, 
Phys. Rev. Lett. {\bf 31}, 1218 (1973).

\bibitem{LMKMBSa03}
P.~Lipavsk{\'y}, K.~Morawetz, J.~Kol{\'a}{\v c}ek, J.~Mare{\v s},
E.~H.~Brandt and M.~Schreiber,
Phys. Rev. B in print

\bibitem{B55}
J.~Bardeen, {\em Theory of Superconductivity}
in Handbuch der Physik, Bd. XV, 274 (Springer Berlin, 1956).

\bibitem{G66}
P.~G.~de~Gennes, 
{\em Superconductivity of Metals and Alloys}, Chap.~VII.3
(Benjamin, NY, 1966).

\bibitem{H75}
K.~M.~Hong, Phys. Rev. B {\bf 12}, 1766 (1975).

\bibitem{KW96}
A.~Kleina and K.~F.~Wojciechowski,
{\em Metal Surface Electron Physics} (Elsevier,
Oxford, 1996).

\bibitem{LKMB02}
P.~Lipavsk{\'y}, J.~Kol{\'a}{\v c}ek, K.~Morawetz
and E.~H.~Brandt,
Phys. Rev. B {\bf 65}, 144511 (2002).

\bibitem{B96}
G.~Blatter {\em et al.},
Phys. Rev. Lett. {\bf 77}, 566 (1996).

\bibitem{K01}
T.~Koyama,
J. Phys. Soc. Jpn. {\bf 70}, 2102 (2001).

\bibitem{MK02}
M.~Machida and T.~Koyama,
Physica C {\bf 378}, 443 (2002).

\bibitem{MK03}
M.~Machida and T.~Koyama,
Physica C {\bf 388}, 659 (2003).

\bibitem{MK03L}
M.~Machida and T.~Koyama,
Phys. Rev. Lett. {\bf 90}, 077003 (2003).

\end{references}
\end{document}